\documentclass[twocolumn,amsmath,amssymb, aps]{revtex4}

\usepackage{graphicx}
\usepackage{dcolumn}
\usepackage{physics}
\usepackage{siunitx}
\usepackage{bbold}
\usepackage{bm,color}
\usepackage[normalem]{ulem}

\begin{document}

\title{Controlling spontaneous emission in inertial and dissipative nematic liquid crystals: the role of critical phenomena}

\author{J. H. Nascimento, F. A. Pinheiro, and M. B. Silva Neto}
\affiliation{%
  Instituto de F\'isica, Universidade Federal do Rio de Janeiro, Caixa Postal 68528, Rio de Janeiro - RJ, Brazil\\
}%

\begin{abstract}
We develop a rigorous, field-theoretical approach to the study of spontaneous emission in inertial and dissipative nematic liquid crystals, disclosing an alternative application of the massive Stueckelberg gauge theory to describe critical phenomena in these systems. This approach allows one not only to unveil the role of phase transitions in the spontaneous emission in liquid crystals but also to make quantitative predictions for quantum emission in realistic nematics of current scientific and technological interest in the field of metamaterials. 
Specifically, we predict that one can switch on and off quantum emission in liquid crystals by varying the temperature in the vicinities of the crystalline-to-nematic phase transition, for both the inertial and dissipative cases. We also predict from first principles the value of the critical exponent that characterizes such a transition, which we show not only to be independent of the inertial or dissipative dynamics,  but also to be in good agreement with experiments. We determine the orientation of the dipole moment of the emitter relative to the nematic director that inhibits spontaneous emission, paving the way to achieve directionality of the emitted radiation, a result that could be applied in tuneable photonic devices such as metasurfaces and tuneable light sources.   
\end{abstract}

\maketitle

\section{\label{sec:level1}INTRODUCTION}

Phase-change materials play a pivotal, increasingly important role in current photonics research. Indeed, they exhibit drastic transitions of their physical and optical properties, sometimes accompanied by internal structural changes, that allow for the development of reconfigurable, tuneable optical devices~\cite{jeong2019}. A typical approach to achieve tuneable optical functionalities consists of incorporating phase-change materials into photonic metasurfaces, planar plasmonic or dielectric nanoresonators that interact with incident light to control the amplitude, phase, wavelength, and polarization of the scattered radiation~\cite{jeong2019}.  For instance this approach harnesses metamaterials for practical applications that require a fast dynamic and programmable optical response, such as active metasurfaces based on phase-changing germanium antimony telluride (GST)~\cite{gst1,gst2,gst3,gst4,gst5,gst6} and vanadium dioxide VO$_{2}$~\cite{vo21,vo22,vo23,vo24,vo25,vo26,vo27,vo28,vo29,vo2110,vo2111}. However, most metasurfaces are passive and hence depend on external light sources to produce the desired optical properties.  

To circumvent this limitation, quantum emitters have been integrated into photonic structures in such a way that the emitted radiation can be tailored by the metasurface design. One of the crucial elements to tailor quantum emission is to engineer Mie-type resonances supported by metasurfaces to enhance light-matter interactions and achieve desired properties such as divergence and directionality~\cite{tanaka2010,langguth2013,staude2015,vaskin2018,liu2018}, sought-after properties in the emerging field of ``smart lighting" and single-photon sources.  

When metasurfaces incorporate both quantum emitters and critical materials, an unprecedented level of control of the emitted radiation can be reached. Indeed, the high sensitivity of the spontaneous emission to the local electromagnetic environment makes the Purcell effect especially prone to be influenced by phase transitions in matter. There exist many examples of drastic modifications of the spontaneous emission rate in critical media, such as a topological phase transition in hyperbolic metamaterials~\cite{Mirmoosa}, percolation transitions in metallic films~\cite{Krachmalnicoff2010} and composite media~\cite{szilard2016}, structural phase transitions in disordered photonic media~\cite{juanjo1,juanjo2}, and structural phase transitions in VO$_{2}$~\cite{szilard2019hysteresis}. All these examples motivated a demonstration that a deeper connection between spontaneous emission and critical phenomena does exist, allowing for probing critical exponents via the Purcell effect~\cite{neto2017characterizing}.

A very recent and promising approach that integrates quantum emitters and critical media into metamaterials to tailor optical functionalities involves the use of Liquid Crystals (LC)~\cite{gorkunov2015metamaterials,Rechcinska2019Engineering}. Nematic LC in particular are interesting for tunable metasurfaces due to their large birefringence and to the fact they undergo structural phase transitions by varying external parameters, such as temperature and applied bias voltage~\cite{gorkunov2015metamaterials,de1993physics}. Exploring these properties of LC, temperature-controlled dynamic beam deflectors~\cite{komar2018dynamic}, tunable silicon metasurfaces~\cite{komar2017electrically}, and controlled light emission using Mie-resonant dielectric metasurfaces~\cite{bohn2018active} have been realized. In these applications understanding the underlying physics of spontaneous emission of emitters embedded in LC is crucial, notably at criticality. Some numerical~\cite{penninck2012light} and theoretical~\cite{mavrogordatos2013spontaneous} studies on spontaneous emission in LC do exist, but to the best of our knowledge none of them investigate the role of LC phase transitions on quantum emission, which is precisely the crucial element for recent applications, as discussed above.

To fill this gap in the present paper we unveil the role of phase transitions in inertial and dissipative nematic LCs in the presence of an embedded quantum emitter (atoms, molecules, or quantum dots). To achieve this goal we develop a rigorous, field-theoretical approach to describe spontaneous emission in nematic liquid crystals. This approach discloses a novel application of the massive Stueckelberg theory, originally developed in the context of a massive but gauge invariant sector of the standard model and with applications to the compactification of high-dimensional string theories \cite{koers2005}, to describe critical phenomena in liquid crystals. It is worth mentioning that other interesting applications of mathemathical physics to LCs and photonics structures do exist \cite{volovik2003universe,longhi2009quantum,timofeev2015geometric,batz2008linear}, but to the best of our knowledge this is the first devoted to quantum emission. In addition to this formal, methodological progress, our theory allows for quantitative predictions for spontaneous emission in nematic liquid crystals used in experiments of current interest in the field of metasurfaces~\cite{bohn2018active}. Specifically we predict a reentrant behaviour of spontaneous emission as a function of temperature near the regime of the crystalline-to-nematic structural phase transitions in nematics, for the case of inertial and dissipative systems. We also show that the spontaneous emission rate can be used to determine the critical exponents that characterize the nematic-to-isotropic transitions in nematics of both scientific and technological interest, even when dissipation becomes dominant. We demonstrate how the relative position between the nematic director and the dipole moment of the emitter may inhibit spontaneous emission in a controlled way, setting the theoretical grounds to achieve directionality of the emitted radiation, which could be incorporated in novel photonic structures.      

This paper is organized as follows: in section \ref{sec:level2} we describe the methodology. We start by recalling the definition of the spontaneous emission rate in free space, given in terms of the spectral function of the electromagnetic gauge field modes. Next we write down a single particle Hamiltonian for an isolated, orientational LC molecule in the presence of the electromagnetic field of a nearby quantum emitter. In \ref{sec:level3} we show that the collective behaviour of all orientational molecules in the uniaxial, nematic LC, and in the presence of the emitter's gauge field is described in terms of a Frank-Oseen free energy density in which the director order parameter, albeit neutral, becomes covariantly coupled to the emitter's vector gauge potential, in complete analogy to Stueckelberg's electromagnetic gauge theory. Once this analogy is established, we conclude by writing down a full, gauge invariant, free energy density for directors in an uniaxial, nematic LC that is nevertheless coupled to the emitter's massive gauge field. In section \ref{sec:level4} we study the critical properties of the problem formulated and solve the quantum action in the saddle-point. In section \ref{sec:level5} we apply the developed theory to make predictions related to quantum emission in realistic liquid crystals of current interest in the field of metamaterials. Finally, section \ref{sec:level6} is devoted to the concluding remarks. 
 
\section{\label{sec:level2}METHODOLOGY}

\subsection{\label{subsec1:level2}Spontaneous emission rate}
Let us begin by considering a quantum emitter whose transition frequency between its ground and excited states is $\omega_{0}$. The spontaneous emission rate of this two-level system in free space is given by \cite{novotny2012principles}
\begin{equation} \label{spontaneous_emission_rate_free_space}
    \Gamma_{0}=\frac{\omega_{0}^{3}|\mathbf{d}|^{2}}{3\pi\varepsilon_{0}\hbar c^{3}}=\frac{\pi\omega_{0}}{3\hbar\varepsilon_{0}}|\mathbf{d}|^{2}\rho_{0}(\omega_{0}),
\end{equation}
where $\rho_{0}(\omega)=\omega^{2}/\pi^{2}c^{3}$ is the electromagnetic density of states in vacuum, given in terms of the speed of light, $c$, $\varepsilon_0$ is the dielectric constant in vacuum, and $\mathbf{d}$ is the electric dipole moment of the emitter. We will now introduce a generalized, bare density of states $g^{2}\rho_0(\omega)$ defined as
\begin{equation} \label{generalized_density_of_states}
    g^{2}\rho_0(\omega)=\frac{1}{\hbar\varepsilon_{0}}\sum_{\lambda=\pm}\!\int\!\frac{d^{3}\mathbf{k}}{(2\pi)^{3}}|\hat{\varepsilon}_{\mathbf{k},\lambda}\cdot\mathbf{d}|^{2}\omega_{\mathbf{k}}^{2}\mathcal{A}^{(0)}(\omega_{\mathbf{k}},\omega),
\end{equation}
where $\hat{\varepsilon}_{\mathbf{k},\lambda}$ is the electromagnetic polarization versor, $\omega_{\mathbf{k}}=c|\mathbf{k}|$ is the photon's dispersion relation, and $\mathcal{A}^{(0)}(\omega_{\mathbf{k}},\omega)$ is the diagonal part of the free-photon spectral function $\mathcal{A}_{\mu\nu}^{(0)}(\omega_{\mathbf{k}},\omega)=\eta_{\mu\nu}\mathcal{A}^{(0)}(\omega_{\mathbf{k}},\omega)$, with \cite{Abrikosov:107441}
\begin{equation} \label{spectral_function_free_photon}
    \mathcal{A}_{\mu\nu}^{(0)}(\omega_{\mathbf{k}},\omega)=\frac{\eta_{\mu\nu}}{2\omega_{\mathbf{k}}}[\delta(\omega-\omega_{\mathbf{k}})-\delta(\omega+\omega_{\mathbf{k}})].
\end{equation}
The above spectral function can be readly obtained from the free-photon propagator in the Feynman gauge \cite{Abrikosov:107441}
\begin{equation} \label{propagator_free_photon}
    \mathcal{G}_{\mu\nu}^{(0)}=\frac{i\eta_{\mu\nu}}{k^{2}}
\end{equation}
with $k=(\omega/c,\mathbf{k})$ and $\eta_{\mu\nu}=\text{diag}(+1,-1,-1,-1)$. The generalized, bare density of states $g^2\rho_0(\omega)$, as defined in Eq. \eqref{generalized_density_of_states}, enables us to accurately include any changes in the environment and/or boundary conditions through its renormalization. As a result, it allows us to rewrite the spontaneous emission rate as
\begin{equation} \label{spontaneous_emission_rate_general}
    \Gamma=\left(\frac{2\pi}{\hbar}\right) g^{2}\rho(\omega),
\end{equation}
which is then to be regarded as a generalization to Fermi's golden rule for interacting, bound systems via the replacement of the generalized bare by the renormalized density of states, $g^2\rho_0(\omega)\rightarrow g^2\rho(\omega)$. In fact,
equation \eqref{spontaneous_emission_rate_general} is a rather general expression and holds also in the case of interacting fields, or nontrivial boundary conditions, after the suitable replacement of the free-photon spectral function $\mathcal{A}^{(0)}(\omega_{\mathbf{k}},\omega)$ by the interacting one $\mathcal{A}(\omega_{\mathbf{k}},\omega)$ in the definition of the generalized, bare $g^2\rho_0(\omega)$ in Eq. \eqref{generalized_density_of_states}.
\newline

\subsection{\label{subsec2:level2}Coupling the director to the field of the emitter}
Liquid crystals may consist of either polar or non-polar molecules. For the case of polar molecules, the permanent electric-dipole moment of the molecules stems from the separation between negative and positive charges within each molecule. We shall denote by $\pmb{\wp}$ such permanent, molecular electric-dipole moment, not to be mistaken with ${\bf d}$, the emitter's electric-dipole moment. The immediate effect of placing an emitter embedded in a liquid crystal is therefore a dipole-dipole interaction described by a Hamiltonian \cite{zangwill2013modern}
\begin{equation} \label{dipole-dipole_interaction_Hamiltonian}
    \mathcal{H}_{\text{int}}\sim\frac{\pmb{\wp}\cdot{\bf d}}{R^3}-\frac{3(\pmb{\wp}\cdot{\bf R})({\bf d}\cdot{\bf R})}{R^5},
\end{equation}
which describes the direct coupling between the emitter, ${\bf d}$, and the LC molecules, $\pmb{\wp}$, separated by a distance ${\bf R}$. Such coupling is naturally expected to enhance spontaneous emission simply by breaking the rotational symmetry of the environment surrounding the emitter, acting effectively as an antenna.

In this work, however, we shall be interested on the effects of the critical fluctuations of the order parameter in the LC to the allowed electromagnetic modes for the emitted radiation. In other words, we shall be interested to investigate how a nematic-isotropic phase transition in a LC can affect the spontaneous emission through changes in the generalized electromagnetic density of states, $g^2\rho(\omega)$, defined in Eq. \eqref{generalized_density_of_states}. For that purpose, we need to describe first how the electromagnetic field couples to the molecular electric-dipole moment of a single molecule in the LC. 

To this end, let us consider a molecular electric-dipole composed by two particles of mass $m$, one with electric charge $-q$ and located at $\mathbf{r}_{1}$ and the other with electric charge $+q$ and located at $\mathbf{r}_{2}$, as a model to the charge separation that occurs inside the molecules of a polar LC. As we have shown in Appendix \ref{SingleMolHamiltonian}, the Hamiltonian for this dipole $\pmb{\wp}$ is given by
\begin{equation} \label{hamiltonian_quantum_emitter}
    \mathcal{H}=\frac{1}{2m\ell^{2}}\left[\mathbf{L}-i\wp\cos\left(\frac{\ell}{2}\mathbf{k}\cdot\mathbf{n}\right)\left(\mathbf{n}\times\mathbf{A}\right)\right]^{2},
\end{equation}
where $m$ is the reduced mass, and $\ell$ is the length of the dipole with dipole moment, $\wp=q\ell$, oriented along $\mathbf{n}$, which points from the negative to the positive charges. Remarkably, a ``Stueckelberg, covariant coupling'' exists in Eq. (\ref{hamiltonian_quantum_emitter}) for the quantized angular momentum
\begin{equation}\label{angular-momentum}
{\bf L}=-\frac{i\hbar}{\ell}\,{\bf n}\times\nabla,    
\end{equation}
which becomes {\it shifted} by $(e^{*}/\hbar)({\bf n}\times {\bf A})$, that is, by the vector product between the director, ${\bf n}$, and the vector potential, ${\bf A}$, through an effective charge $e^{*}$ given by
\begin{equation} \label{effective_charge}
    e^{*}=\frac{2\wp}{\ell}\cos\left(\frac{\ell}{2}\mathbf{k}\cdot\mathbf{n}\right),
\end{equation}
and this is going to be explored below. This Stueckelberg coupling is the reason why we considered only the rotational degree of freedom of Eq. \eqref{dipole_hamiltonian_final_form} in Appendix \ref{SingleMolHamiltonian}.

\subsection{\label{subsec3:level2} Uniaxial nematic LC coupled to an electromagnetic field}
The most general continuum field theoretical description for nematic liquid crystals is the Landau$-$de-Gennes (LdG) theory \cite{de1993physics}. According to the LdG theory, for uniaxial LCs the local order parameter can be written as a traceless, rank$-2$ tensor, $Q_{ij}({\bf r})$, with Cartesian indices $i,j=1,2,3$, that can also be written as \cite{de1993physics}
\begin{equation}
    {\bf Q}({\bf r})=s\left({\bf n}({\bf r})\otimes {\bf n}({\bf r})-\frac{\mathbb{I}}{3}d\right),
\end{equation}
where $\mathbb{I}$ is the identity tensor in $d$ dimensions, the order parameter magnitude, $s$, determines the strength of the orientational ordering of the nematic molecules, while the symmetry axis is determined by the eigenvector ${\bf n}$ of ${\bf Q}({\bf r})$, with ${\bf n}^2=1$. In order to include spatial variations of such order parameter one usually writes down a LdG free energy \cite{de1993physics}
\begin{equation}
    \mathcal{F}_{\text{LdG}}[{\bf Q}]=\int\!d^{3}\mathbf{r}\; \left\{\frac{L}{2}\left|\nabla {\bf Q}({\bf r})\right|^2+f_{LG}({\bf Q}({\bf r}))\right\}
\end{equation}
where the Landau-Ginzburg part is given by \cite{de1993physics}
\begin{equation}
    f_{LG}({\bf Q}({\bf r}))=\frac{\alpha(T-T^*)}{2}\text{tr}({\bf Q}^2)-\frac{b^2}{3}\text{tr}({\bf Q}^3)+\frac{c^2}{4}\text{tr}({\bf Q}^2)^2,
\end{equation}
where $\alpha(T-T^*), b^2,$ and $c^2$ are material-dependent and temperature-dependent nonzero constants, while $L$ is the elastic constant. In terms of free energy ${\cal F}_{LdG}[{\bf Q}]$ the LC is said to be in the {\it isotropic regime} for $T>T^*$, when $\alpha(T-T^*)>0$, and the free energy is {\it globally minimized} for ${\bf Q}=0$ \cite{Majumdar2010}. In the {\it nematic ordered regime}, in turn, for $T<T^*$, when $\alpha(T-T^*)=-a^2<0$, the free energy ${\cal F}_{LdG}[{\bf Q}]$ is {\it globally minimized} for finite values of the order parameter, ${\bf Q}^{(L)}\neq 0$, which depend on the elastic constant $L$ \cite{Majumdar2010}. 

For uniaxial, nematic LCs the elastic constant $L$ is typically very small, $L\sim 10^{-11}J/m$, and for this reason one usually considers the $L\rightarrow 0$ limit \cite{Majumdar2010}. This is the limit we shall be considering in the present work, the so called {\it limiting harmonic map}, and in this case, the order parameter configuration that minimizes the LdG free energy is reduced to \cite{Majumdar2010}
\begin{equation}
    {\bf Q}^{(0)}({\bf r})=s_+\left({\bf n}^{(0)}({\bf r})\otimes {\bf n}^{(0)}({\bf r})-\frac{\mathbb{I}}{3}d\right),
\end{equation}
where $s_+=(b^2+\sqrt{b^4+24a^2c^2})/4c^2$ and ${\bf n}^{(0)}$ is the minimizing director configuration of the Frank-Oseen free energy per unit of volume \cite{de1993physics,Majumdar2010}
\begin{equation}
    f_d({\bf n})=\frac{1}{2}K_{1}(\nabla\cdot\mathbf{n})^{2}+\frac{1}{2}K_{2}(\mathbf{n}\cdot\nabla\times\mathbf{n})^{2}+\frac{1}{2}K_{3}(\mathbf{n}\times\nabla\times\mathbf{n})^{2}.
\end{equation}
In the above expression the elastic constants $K_{i=1,2,3}$, measure the resistance to the three simplest types of distortion: splay, twist, and bend, and we have dropped the superscript $(0)$ to the director order parameter ${\bf n}$. For the sake of simplicity, but without loss of generality, we shall further consider only bend distortions in the uniaxial, nematic system so that $K_{1}=K_{2}=0$ and 
\begin{align} \label{frank-oseen_free_energy_bend}
    f_{\text{bend}}({\bf n})=\frac{1}{2}K_{3}(\mathbf{n}\times\nabla\times\mathbf{n})^{2}=\frac{1}{2}K_{3}\left[\left(\mathbf{n}\cdot\nabla\right)\mathbf{n}\right]^{2}.
\end{align}
This approximation is not only realistic but of practical interest since there exist nematic LC's with only one kind of distortion shown in \eqref{frank-oseen_free_energy_bend}, as remarked in \cite{de1993physics}. 

Let us now take a closer look at the Eq. \eqref{hamiltonian_quantum_emitter}. Considering only the bend distortion, and considering the definition of the angular momentum operator, ${\bf L}$, given in Eq. (\ref{angular-momentum}), one concludes that Eq.\eqref{hamiltonian_quantum_emitter} leads to a shift of the gradient operator, or equivalently, to a Stueckelberg-type \cite{stuckelberg1938wechselwirkungskrafte,stueckelberg1938forces}, covariant coupling prescription between the director of the nematic and the emitted photon, through the effective charge $e^{*}$ defined in Eq. (\ref{effective_charge}). By adopting such prescription we are allowed to write $\nabla\rightarrow\nabla-(e^{*}/\hbar)\mathbf{A}$ in Eq. \eqref{frank-oseen_free_energy_bend}, which gives
\begin{equation} \label{frank-oseen_free_energy_bend_minimal_coupling}
    f_{\text{bend}}^{\text{EM}}({\bf n})=\frac{1}{2}K_{3}\left\{(\mathbf{n}\cdot\nabla)\mathbf{n}+\frac{e^{*}}{\hbar}\left[\mathbf{A}-(\mathbf{n}\cdot\mathbf{A})\mathbf{n}\right]\right\}^{2}.
\end{equation}
From Eq. \eqref{effective_charge} we conclude that with respect to the nematic order, described by $\mathbf{n}$, the preferential direction of propagation is perpendicular to it, so we have $\mathbf{k}\perp\mathbf{n}$ and $e^{*}$ is maximal.

\section{\label{sec:level3} Stueckelberg Gauge theory for the uniaxial nematic LC}

From now on, we will adopt a more suitable notation where we decompose the director, $\mathbf{n}$, of the LC according to $\mathbf{n}=\pmb{\sigma}+\pmb{\pi}$, with $\pmb{\sigma}$ and $\pmb{\pi}$ describing its equilibrium and transverse fluctuations, respectively, $\pmb{\sigma}\perp\pmb{\pi}$. In this notation we can write each component of $\mathbf{n}$ as $n_{i}=\sigma_{i}+\pi_{i}$. Furthermore, we shall also note that, in the absence of the electromagnetic field, the Lagrangian density describing the fluctuations, $\pmb{\pi}$, of the otherwise homogeneous, $\pmb{\sigma}_0$, nematic order is given simply by the distortion free energy itself, given in \eqref{frank-oseen_free_energy_bend}, yielding
\begin{equation} \label{lagrangean_fluctuations_nematic_order_absence_EM_field}
    \mathcal{L}_{\text{bend}}=\frac{1}{2}K_{ii'}(\partial_{i}\pi_{j})(\partial_{i'}\pi_{j})
\end{equation}
with $K_{ii'}=K_{3}\sigma_{i}\sigma_{i'}$ and a summation notation is implied. From \eqref{lagrangean_fluctuations_nematic_order_absence_EM_field}, the propagator of the fluctuations reads \cite{Abrikosov:107441}
\begin{equation} \label{propagator_nematic_bend_absence_EM_field}
    G_{ii'}(\mathbf{p})=\frac{1}{K_{ii'}p_{i}p_{i'}}.
\end{equation}
Moving on to the case where we have the coupling to the electromagnetic field, the Lagrangian is now 
\begin{equation} \label{lagrangean_fluctuations_nematic_order_presence_EM_field}
    \mathcal{L}_{\text{bend}}^{\text{EM}}=\frac{1}{2}K(\partial_{i}\pi_{j})^{2}+\frac{1}{2}JA_{j}(\partial_{i}\pi_{j})+\frac{1}{2}M^{2}A_{k}^{2},
\end{equation}
where we have made the consideration $\sigma_{i}\sigma_{j}=\Lambda\delta_{ij}$, for all $i,j$ with $\Lambda$ being a function of temperature (whose analytical form will be given later on in the text) and we have also defined
\begin{gather} \label{parameters}
    K=K_{3}\Lambda \\
    M^{2}=\frac{e^{*2}}{\hbar^{2}}K_{3}(1-\Lambda) \\
    J=2\frac{K_{3}e^{*}}{\hbar}\sqrt{\Lambda}(1-\Lambda).
\end{gather}

Using natural units $\hbar=c_s=k_B=1$ and considering fluctuations, $\pmb{\pi}$, on top of a homogeneous, equilibrium configuration, $\pmb{\sigma}$, we end up with a gauge Lagrangian density describing the director fluctuations covariantly coupled to the transverse components of the emitter's electromagnetic vector potential
\begin{equation} \label{quantum_action_bend_fluctuations_minimal_coupling}
    \mathcal{L}_{\text{bend}}^{\text{gauge}}=\frac{1}{2}(\nabla\times {\bf A}_\perp)^2
    +\frac{1}{2}K_3[(\pmb{\pi}\cdot\nabla)\pmb{\pi}+e^{*}\mathbf{A}_{\perp}]^{2},
\end{equation}
where $\mathbf{A}_{\perp}=\mathbf{A}-(\pmb{\sigma}\cdot\mathbf{A})\pmb{\sigma}$ are the components of the vector potential transverse to the order parameter, $\pmb{\sigma}$. Recalling the definition of the effective charge $e^{*}$ given in \eqref{effective_charge}, we see that the preferred direction of propagation is the one perpendicular to the nematic order parameter ${\bf n}\parallel\pmb{\sigma}$, since $\cos\left(\frac{\ell}{2}\mathbf{k}\cdot\mathbf{n}\right)=1$ and $e^{*}$ is maximum, while the photon remains massless when transverse to $\pmb{\sigma}$, since in this case $\pmb{\sigma}\cdot\mathbf{A}_{\perp}=\pmb{\sigma}\cdot(\mathbf{A}-(\pmb{\sigma}\cdot\mathbf{A})\pmb{\sigma})=0$. The three vectors, ${\bf k},\;\pmb{\sigma},$ and ${\bf A}$ define completely the possibilities of electromagnetic wave propagation and polarizability inside the LC, which are encoded in $e^*$, Eq. \eqref{effective_charge}. 

In order to show that our theory is gauge invariant, despite being massive in the gauge sector, we begin by considering the following gauge transformations
\begin{equation} \label{gauge_transformation_field_A}
    \mathbf{A}_{\perp}\rightarrow\mathbf{A}'_{\perp}=\mathbf{A}_{\perp}-(\pmb{\sigma}\cdot\nabla)\pmb{\Psi}
\end{equation}
and
\begin{equation} \label{gauge_transformation_field_pi}
    \pmb{\pi}\rightarrow\pmb{\pi}'=\pmb{\pi}+e^{*}\pmb{\Psi},
\end{equation}
where $\pmb{\Psi}$ is a gauge function valid for a real and neutral field such as $\pmb{\pi}$ field. The gauge invariance is shown explicitly as follows:
\begin{center}
\textit{(i) covariant derivative term}
\end{center}
\begin{align}
    (\pmb{\sigma}\cdot\nabla)\pmb{\pi}'+e^{*}\mathbf{A}'_{\perp}&=(\pmb{\sigma}\cdot\nabla)(\pmb{\pi}+e^{*}\pmb{\Psi})\;+\nonumber\\
    &\qquad\qquad\qquad+e^{*}(\mathbf{A}_{\perp}-(\pmb{\sigma}\cdot\nabla)\pmb{\Psi})\nonumber\\
    &=(\pmb{\sigma}\cdot\nabla)\pmb{\pi}+e^{*}\mathbf{A}_{\perp},
\end{align}
\begin{center}
\textit{(ii) gauge field term}
\end{center}
\begin{align}
    (\nabla\times\mathbf{A}'_{\perp})^{2}&=\left\{\nabla\times[\mathbf{A}_{\perp}-(\pmb{\sigma}\cdot\nabla)\pmb{\Psi}]\right\}^{2}\nonumber\\
    &=(\nabla\times\mathbf{A}_{\perp})^{2}+(\nabla\times(\pmb{\sigma}\cdot\nabla)\pmb{\Psi})^{2}\;+\nonumber\\
    &\qquad\qquad-2(\nabla\times\mathbf{A}_{\perp})\cdot(\nabla\times(\pmb{\sigma}\cdot\nabla)\pmb{\Psi})\nonumber\\
    &=(\nabla\times\mathbf{A}_{\perp})^{2},
\end{align}
as long as the gauge function $\pmb{\Psi}$ satisfies the condition
\begin{equation} \label{condition_over_Psi}
    \nabla\times(\pmb{\sigma}\cdot\nabla)\pmb{\Psi}=0.
\end{equation}

From \eqref{lagrangean_fluctuations_nematic_order_presence_EM_field} it is clear that our problem possesses some similarities with the massive Stueckelberg gauge theory, due to a mass term $M^{2}A_{k}^{2}$ and gauge invariance. For this reason, we believe is worth giving a brief review about this theory. 

Stueckelberg's theory deals with an Abelian massive vector field $A_{\mu}$ as well as a massive scalar field $B$, the latter called Stueckelberg field, coupled to the $A_{\mu}$-field. The $B$-field is required to properly quantize this theory \cite{stuckelberg1938wechselwirkungskrafte,stueckelberg1938forces}. These fields have the same mass and we will denote it by $\mathcal{M}$ to avoid confusion with our mass term $M$ previously defined. The Stueckelberg Lagrangian is
\begin{multline} \label{stueckelberg_massive_lagrangian}
    \mathcal{L}=-\frac{1}{4}F_{\mu\nu}F^{\mu\nu}+\frac{1}{2}\mathcal{M}^{2}\left(A^{\mu}-\frac{1}{\mathcal{M}}\partial^{\mu}B\right)^{2}-\\
    -\frac{1}{2\xi}(\partial_{\mu}A^{\mu}+\xi\mathcal{M}B)^{2},
\end{multline}
where the third term is the gauge fixing term, and this Lagrangian has the advantage of being manifestly covariant and gauge invariant even for a massive vector field, in contrast to the Proca Lagrangian \cite{ryder1996quantum}. For details on the well-established Stueckelberg theory, we refer to Ref. \cite{ruegg2004stueckelberg}.

In fact, by considering a gauge fixing term and keeping the terms up to second order in the gauge field, the full Lagrangian of our problem can be written as
\begin{equation} \label{full_lagrangean_of_our_problem}
    \mathcal{L}=-\frac{1}{4}F_{\mu\nu}F^{\mu\nu}+\frac{1}{2}M^{2}A_{\mu}A^{\mu}-\frac{1}{2\xi}(\partial_{\mu}A^{\mu})^{2}.
\end{equation}
It is important to keep in mind that although we use a relativistic notation in \eqref{full_lagrangean_of_our_problem}, our problem lies in the Euclidian space and, hence, we are actually working with three-vectors. In passing from one notation to the other, we have used the gauge choice $A^{0}=A_{0}=0$. Working in the Feynman gauge, the propagator associated with \eqref{full_lagrangean_of_our_problem} is given by \cite{Abrikosov:107441}
\begin{equation} \label{propagator_of_our_problem}
    \mathcal{G}_{ij}^{M}(\mathbf{k},\omega)=\frac{\delta_{ij}}{\omega^{2}-(\mathbf{k}^{2}+M^{2})},
\end{equation}
where we came back to the Latin indices associated with the Euclidian space.

The comparison between our model to the original Stueckelberg theory becomes evident by noting that (i) the original neutral and scalar Stueckelberg field $B$ defined in \cite{ruegg2004stueckelberg} is replaced, in our theory, by a neutral and vector field $\pmb{\pi}$, (ii) the gauge function $\pmb{\Psi}$ in the Stueckelberg theory satisfies $(\nabla^{2}+m^{2})\pmb{\Psi}=0$ while in our model it satisfies $\nabla\times(\pmb{\sigma}\cdot\nabla)\pmb{\Psi}=0$, and (iii) the Stueckelberg field $B$ is massive while our $\pmb{\pi}$ is massless, as it should be in accordance with Goldstone’s theorem \cite{Peskin:257493}. We hence disclose for the first time another application for the massive Stueckelberg theory to describe quantum
emission in LC's.

\section{\label{subsec1:level4}Nematodynamics of the director}

Before calculating the spontaneous emission rate, let us recall that the gauge mass, $M$, of our St\"uckelberg model for quantum emitters embedded in a nematic LC depends on the nematic order parameter, $\Lambda=\sigma^2$, via Eq. \eqref{parameters}. In order to calculate such mass dependence self-consistently, we must first and foremost obtain the evolution of the order parameter, $\sigma$, including quantum and thermal fluctuations of the director. 

\subsection{\label{sec:level4}Porod tails and the $O(N)$-nonlinear $\sigma$ model}
We first note that the Landau-De Gennes theory for uniaxial nematic LCs has a strong analogy to the 3D version of the Ginzburg-Landau theory for superconductors \cite{Bethuel1993}. For the case of our three dimensional director field ${\bf n}:\Omega\rightarrow {\cal R}^3$ the appropriate Ginzburg-Landau gauge energy functional is of the form
\begin{multline}
    {\cal F}_{GL}[{\bf A}_\perp,{\bf n}]=\int_\Omega\, d^3{\bf r}\left\{\frac{1}{2}K_3(\mathbf{n}\cdot\nabla)\mathbf{n}+\frac{1}{4\epsilon^2}({\bf n}^2-1)^2\right\}\nonumber\\
    + f_{GL}({\bf A}_\perp,{\bf n}),
\end{multline}
where $f_{GL}({\bf A}_\perp,{\bf n})$ includes pure gauge and gauge interaction contributions to the free energy functional. This functional has been rigorously studied in the $\epsilon\rightarrow 0$ limit, which corresponds to the {\it limiting harmonic map}, $L\rightarrow 0$ limit, of our problem \cite{Majumdar2010,Bethuel1993}. The quartic, $({\bf n}^2-1)^2$, interaction term in the $O(N)$-vector model enforces a {\it soft} constraint to the director and thus allows for both transverse as well as longitudinal fluctuations of the order parameter, which are specially relevant near criticality \cite{Bochkarev1996}. The {\it hard}, fixed length constraint, $\delta({\bf n}^2-1)$, instead, only allows for transverse fluctuations \cite{Bochkarev1996}. Since at large wavevectors the nematic structure factor $S(k)$ is dominated by power-law contributions originating from singular variations associated to both topological defects and transverse thermal fluctuations of the nematic order parameter, the so called Porod tails, $S(k)\sim k^{-\xi}$ \cite{Bray2002}, where $\xi$ is the Porod exponent, in what follows we shall use the fixed length constraint, $\delta({\bf n}^2-1)$, clearly justified in the {\it limiting harmonic map}, $L\rightarrow 0$ limit, and the large $N$ limit.

\subsection{\label{subsec2:level4}Ericksen-Leslie dynamics - inertia vs dissipation}

Next we must introduce the proper dynamics for the director fluctuations, $\mathbf{n}$. We shall do this by using Ericksen-Leslie (EL) equations for the dynamics of inertial and dissipative nematic liquid crystals, which are widely accepted and have been experimentally verified extensively \cite{Ericksen1962,Leslie1968}. Initially we focus on inertial effects only and shall thus neglect the effects of dissipation. These will be included subsequently and compared to the results for the inertial case.

The dynamics of the director, ${\bf n}$, is considered to be governed by Ericksen-Leslie equations \cite{Gay-Balmaz2013,Gay-Balmaz2012}
\begin{equation}
    I\frac{\partial^2{\bf n}}{\partial t^2}-\left[{\bf n}\cdot{\bf h}+I{\bf n}\cdot\frac{\partial^2{\bf n}}{\partial t^2}\right]{\bf n}+{\bf h}=0,
\end{equation}
where $I$ is the microinertia constant \cite{Gay-Balmaz2013,Gay-Balmaz2012} and 
\begin{equation}
    {\bf h}=\frac{\partial f_\text{bend}}{\partial {\bf n}}-\frac{\partial}{\partial x^i}\frac{\partial f_\text{bend}}{\partial(\partial_{x^2}{\bf n})},
\end{equation}
is the molecular field expressed in terms of the Frank-Oseen free energy density, $f_\text{bend}$ \cite{Gay-Balmaz2013,Gay-Balmaz2012}.

One can verify that the above EL equation is simply the Euler-Lagrangian equation that arises from the Lagrangian \cite{Gay-Balmaz2013,Gay-Balmaz2012}
\begin{equation}
    \mathcal{L}({\bf n},\partial_t{\bf n},\nabla{\bf n})=\frac{I}{2}(\partial_t{\bf n})^2-f_\text{bend}({\bf n},\nabla{\bf n}),
\end{equation}
which, in turn, produces a dynamical Matsubara propagator \cite{Mahan}, for the transverse fluctuations of the director field, $\pmb{\pi}$, given by
\begin{equation}
    G({\bf p},\omega_n)=\frac{1}{c_0^2\omega_n^2+(\pmb{\sigma}\cdot{\bf p})^2+\xi^{-2}},
\end{equation}
where $c_0\sim I$ is a group velocity that warrants a purely oscillatory dynamics for the problem and $\xi$ is the correlation length that diverges at the isotropic-to-nematic phase transition. As we can see, inertia produces a hyperbolic (linear) dispersion relation for the director fluctuations in the isotropic (nematic) phase
\begin{equation}
    \omega({\bf p})=\frac{1}{c_0}\sqrt{(\pmb{\sigma}\cdot{\bf p})^2+\Delta^2},
\end{equation}
with $\Delta=(c_0 \xi)^{-1}\rightarrow 0$ for $T<T_{NI}$, where $T_{NI}$ is the nematic-isotropic transition temperature.

When dissipation is present the solution is modified from hyberpolic to parabolic \cite{Gay-Balmaz2013,Gay-Balmaz2012}, the reason being the replacement of the inertia by the dissipative term \cite{Gay-Balmaz2013,Gay-Balmaz2012}
\begin{equation}
    I\, D_t^2{\bf n}\rightarrow\gamma \, D_t{\bf n},
\end{equation}
where $D_t=\partial_t+{\bf n}\cdot{\nabla}$ is the material derivative and $\gamma$ is the damping coefficient \cite{Gay-Balmaz2013,Gay-Balmaz2012}. When dissipation becomes dominant, $\gamma\gg I$, it produces a dynamical Matsubara propagator \cite{Mahan} for the transverse director fluctuations, $\pmb{\pi}$, given by
\begin{equation}
    G({\bf p},\omega_n)=\frac{1}{\frac{|\omega_n|}{D}+(\pmb{\sigma}\cdot{\bf p})^2+\xi^{-2}},
\end{equation}
where $D\sim 1/\gamma$ is a diffusion constant that warrants a purely relaxional dynamics for the problem. As we can see, dissipation produces a parabolic (quadratic) dispersion relation for the director fluctuations in the isotropic (nematic) phase
\begin{equation}
    \omega({\bf p})= D (\pmb{\sigma}\cdot{\bf p})^2+\Delta^2,
\end{equation}
with $\Delta=(\sqrt{D}/\xi)\rightarrow 0$ for $T<T_{NI}$.

\subsection{\label{subsec3:level4}Thermal evolution of the order parameter}
We are finally ready to calculate the critical exponents associated to the nematic-to-isotropic transition due to thermal fluctuations of the bosonic order parameter ${\bf n}$ with $N$ components. As discussed earlier, we therefore separate longitudinal, $\pmb{\sigma}$, from transverse, $\pmb{\pi}$, components. The self consistent equation that determines the amplitude of the longitudinal component is \cite{Bochkarev1996}
\begin{equation}
    \pmb{\sigma}^{2}=1-\left\langle \pmb{\pi}\cdot\pmb{\pi}\right\rangle.
\end{equation}
This is simply the equation for the constraint imposed on average at large $N$. Using units where $\hbar=c_0=1$ we will now generalize the inertia  and dissipative cases discussed above by using a phenomenological, Matsubara propagator \cite{Mahan} for the transverse fluctuations of the director field, $\pmb{\pi}$, given by
\begin{equation}
    G({\bf p},\omega_n)=\frac{1}{D^{1-z}|\omega_n|^{2/z}+(\pmb{\sigma}\cdot{\bf p})^2+\xi^{-2}},
\end{equation}
where $1\leq z\leq 2$ is a dynamical exponent that interpolates between the purely oscillatory, $z=1$, and the purely dissipative, $z=2$, cases. Now, when written explicitly for the bosonic transverse fluctuations, the fixed length constraint reduces to \cite{Bochkarev1996}
\begin{equation}
        \pmb{\sigma}^2=1-\int\frac{d^{3}\mathbf{p}}{\left(2\pi\right)^{3}}\frac{g D^{z(1-z)/2}}{|\pmb{\sigma}\cdot\mathbf{p}|^z}\coth\left(\frac{|\pmb{\sigma}\cdot\mathbf{p}|^z}{2 D^{z(1-z)/2} k_B T}\right),
\end{equation}
where $g$ is a dimensionless coupling constant of the order $N$, $k_B$ is the Boltzmann constant, and $T$ is the temperature.
The momentum integration is divergent in the ultraviolet as $\mathbf{p}\rightarrow\infty$. In order to remedy this problem, we introduce the critical temperature $T_{C}$ at which the order parameter vanishes, $\sigma(T_{C})=0$ \cite{Mahan}. In Section \ref{sec:level5}, we shall identify this critical temperature with the nematic-isotropic transition temperature $T_{NI}$. We first introduce the averages
\begin{equation}
    \left|\pmb{\sigma}\cdot\mathbf{p}\right|\rightarrow |\sigma\mathbf{p}|.
\end{equation} 
in terms of which the critical temperature must satisfy
\begin{equation}
    1=\int\frac{d^{3}\mathbf{p}}{\left(2\pi\right)^{3}}\frac{g D^{z(1-z)/2}}{\sigma^z\left|\mathbf{p}\right|^z}\coth\left(\frac{\sigma^z\left|\mathbf{p}\right|^z}{2 D^{z(1-z)/2} k_B T_{C}}\right).
\end{equation}
Now the self-consistent equation for the order parameter reduces to
\begin{multline} \label{saddle-point_equation_for_the_order_parameter}
    \sigma^{2+z}=\frac{g D^{z(1-z)/2}}{2\pi^{2}}\int_{0}^{\infty}dp\;p^{2-z}\times\\
    \times\Bigg[\coth\left(\frac{\sigma^z p^z}{2 D^{z(1-z)/2} k_B T_{C}}\right)-\\
    -\coth\left(\frac{\sigma^z p^z}{2 D^{z(1-z)/2} k_B T}\right)\Bigg],
\end{multline}
which is finite in the ultraviolet and clearly vanishes as $T\rightarrow T_{C}$. We now re-scale momenta
\begin{equation}
    p=(2k_B T_C)^{1/z}\, q\, \frac{D^{(1-z)/2}}{\sigma},
\end{equation}
such that \eqref{saddle-point_equation_for_the_order_parameter} becomes
\begin{multline} \label{saddle-point_equation_for_the_order_parameter_final_form}
    \sigma^5=\frac{g D^{3(1-z)/2}(2 k_B T_{C})^{(3-z)/z}}{2\pi^{2}}\times\\
    \times\int_{0}^{\infty}dq\;q^{2-z}\left[\coth\left(q^z\right)-\coth\left(\frac{q^z}{t}\right)\right],
\end{multline}
where we have defined the reduced temperature $t$ as
\begin{equation} \label{reduced_temperature}
    t=\frac{T}{T_{C}}\leq1.
\end{equation}
If we would like now to calculate the critical exponent of the order parameter close to criticality, we must expand the expression for $\sigma$ close to $t\rightarrow 1$ to obtain
\begin{equation}
    \sigma \approx \sigma_C\left(1-t\right)^{1/5}, 
\end{equation}
where the prefactor is 
\begin{equation}
    \sigma_C = \left\{\frac{g D^{3(1-z)/2}(2 k_B T_{C})^{(3-z)/z}}{2\pi^{2}}\mathcal{I}\right\}^{1/5}
\end{equation}
with $\mathcal{I}=\int_{0}^{\infty}dq\;q^{2}\left[1-\coth^{2}\left(q^z\right)\right]$,
thus producing an exponent
\begin{equation} \label{exponent}
    \beta=1/5.
\end{equation}

Remarkably, this value for the critical exponent, predicted by our model, {\it does not} depend on the inertial versus dissipative dynamics, unveiling its static, high-temperature character, and is in good agreement with experimentally determined values in nematic LC's, such as $\beta=0.241$ \cite{chirtoc2004determination}, $\beta=0.28$ \cite{lenart2012tricritical} and $\beta=0.219$ \cite{haller1975thermodynamic}.

\section{\label{sec:level5} SPONTANEOUS EMISSION RATE IN NEMATIC LIQUID CRYSTALS}

Let us now calculate the spontaneous emission rate for a quantum emitter embedded in a nematic quantum LC. For that, let us observe that the pole of the propagator \eqref{propagator_of_our_problem} lies within the frequency range  $\omega_{\mathbf{k},M}=\pm\sqrt{\mathbf{k}^{2}+M^{2}}$ and it follows that the corresponding spectral function is given by \cite{Abrikosov:107441,Mahan}
\begin{equation} \label{spectral_function_of_our_problem}
    \mathcal{A}_{ij}^{M}(\omega_{\mathbf{k},M},\omega)=\frac{\delta_{ij}}{2\omega_{\mathbf{k},M}}[\delta(\omega-\omega_{\mathbf{k},M})-\delta(\omega+\omega_{\mathbf{k},M})].
\end{equation}
By substituting the diagonal part of Eq. \eqref{spectral_function_of_our_problem} into Eq. \eqref{spontaneous_emission_rate_general}, the spontaneous emission rate as a function of the reduced temperature reads 
\begin{equation} \label{change_in_spontaneous_emission_rate_quantum_emitter}
    \frac{\Gamma(t)}{\Gamma_{0}}=\sqrt{1-\left[\frac{M(t)c^{2}}{\hbar\omega_{0}}\right]^{2}}=\sqrt{1-M_{0}\frac{M^{2}(t)}{M^{2}(t')}},
\end{equation}\\
where $M^{2}(t)=\frac{\overline{e^{*2}}}{\hbar^{2}}K_{3}(t)\Lambda(t)$ is the mass of the gauge field, $M^{2}(t')$ is the mass term calculated at the temperature $t'$ defined as the ratio between the crystalline-nematic and nematic-isotropic transition temperatures and $M_{0}=\frac{M^{2}(t')c^{4}}{\hbar^{2}\omega^{2}_{0}}$ is a positive constant. Here we have considered a spatial average $\overline{e^{*2}}$ over the effective charge \eqref{effective_charge} since the emission of the photon is isotropic inside the nematic. As shown in Eq. \eqref{change_in_spontaneous_emission_rate_quantum_emitter}, the elastic constant $K_{3}$ and $\Lambda$ are both functions of the temperature. For the former we refer to experimental data \cite{bradshaw1985frank,de1976determination,hakemi1983temperature} in order to obtain an analytical form for $K_{3}(t)$ given by $K_{3}(t)=a_{0}-a_{1}t+a_{2}t^{2}-a_{3}t^{3}$,
where $a_{0},...,a_{3}$ are positive constants characterizing different nematic LC's, and for the latter we assume for simplicity, but without loss of generality, the power law $\Lambda(t)=|1-t|^{\beta}$ where $\beta$ is the critical exponent. Thus, by substituting the functional forms of $K_{3}(t)$ and $\Lambda(t)$ into \eqref{change_in_spontaneous_emission_rate_quantum_emitter}, we obtain
\begin{widetext}
    \begin{equation}\label{change_in_spontaneous_emission_rate_quantum_emitter_final_form}
        \frac{\Gamma(t)}{\Gamma_{0}}=\sqrt{1-M_{0}\frac{(a_{0}-a_{1}t+a_{2}t^{2}-a_{3}t^{3})(1-|1-t|^{\beta})}{(a_{0}-a_{1}t'+a_{2}t^{'2}-a_{3}t^{'3})(1-|1-t'|^{\beta})}}.
    \end{equation}
\end{widetext}

\begin{figure}[ht]
    \centering
 	\includegraphics[width=0.49\textwidth]{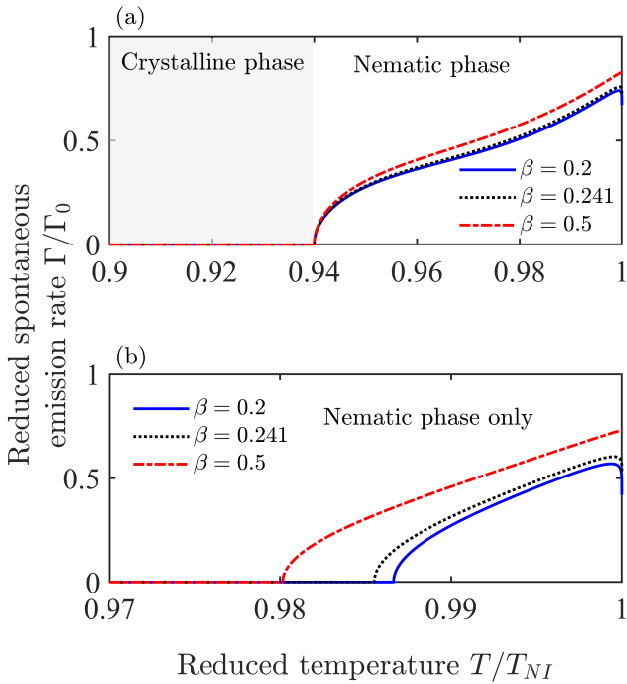}
	 \caption{Spontaneous emission rate $\Gamma$ for 5CB as a function of the reduced temperature $T/T_{NI}$, where $T_{NI}$ is the nematic-isotropic transition temperature. The values of $\Gamma$ have been normalized by its vacuum value $\Gamma_0$. We have set $M_{0}=1$ for panel $(a)$ and $M_{0}=1.5$ for panel $(b)$.}
	\label{spontaneous_emission_5CB_fig}
\end{figure}

\begin{figure}[ht]
    \centering
 	\includegraphics[width=0.49\textwidth]{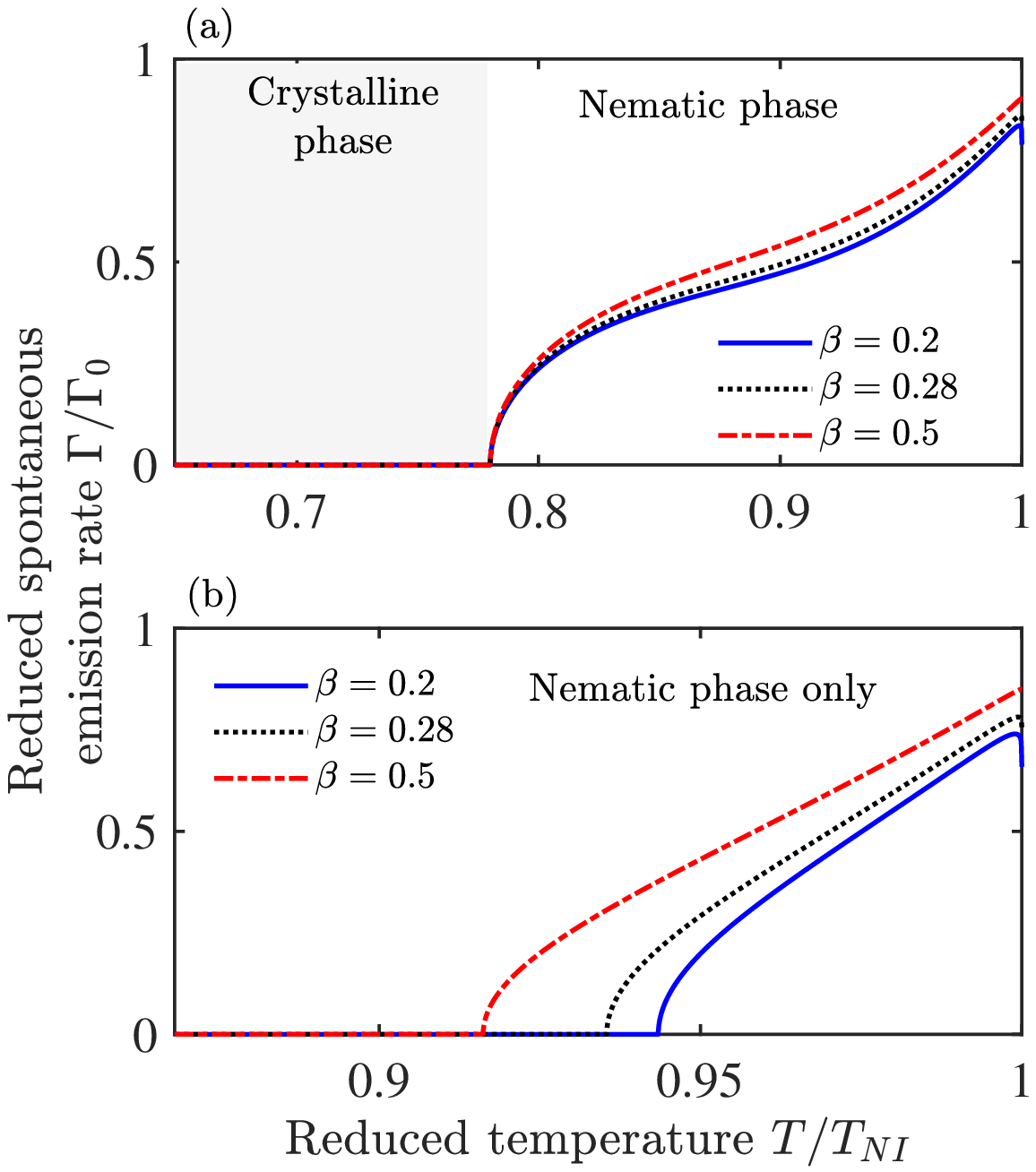}
	 \caption{Spontaneous emission rate for E7 as a function of the reduced temperature $T/T_{NI}$. We have set $M_{0}=1$ for panel $(a)$ and $M_{0}=1.5$ for panel $(b)$.}
	\label{spontaneous_emission_E7_fig}
\end{figure}

\begin{figure}[ht]
    \centering
 	\includegraphics[width=0.49\textwidth]{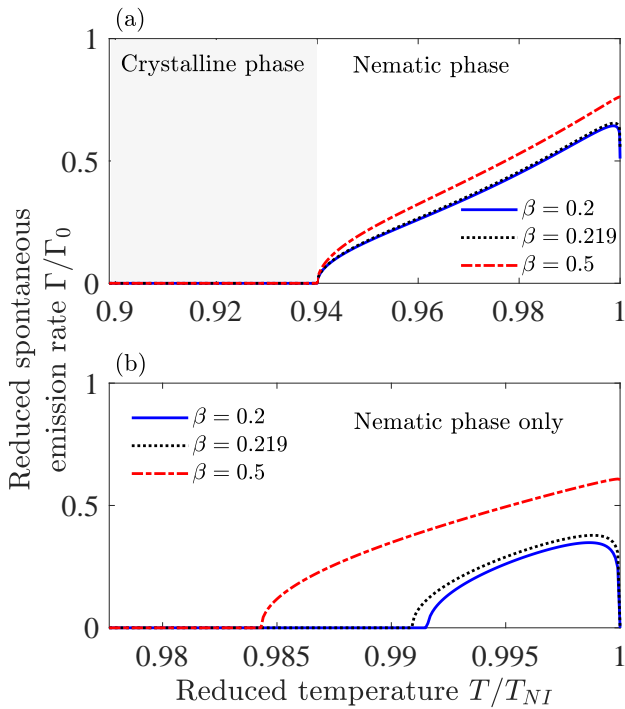}
	 \caption{Spontaneous emission rate for MBBA as a function of the reduced temperature $T/T_{NI}$. We have set $M_{0}=1$ for panel $(a)$ and $M_{0}=1.5$ for panel $(b)$.}
	\label{spontaneous_emission_MBBA_fig}
\end{figure}

Equation \eqref{change_in_spontaneous_emission_rate_quantum_emitter_final_form} gives, to the best of our knowledge, the first analytical expression for spontaneous emission rate $\Gamma$ of a quantum emitter embedded in nematic liquid crystals, valid near the critical point. For concreteness, we consider three different nematic liquid crystals of scientific and technological importance, namely 4-pentyl-4-cyanobiphenyl (5CB), N-(p-methoxybenzylidene)-p-n-butylaniline (MBBA) and the nematic mixture E7, which exhibit nematic to isotropic phase transition at 308.75 K, 318.15 K, 335.65 K, respectively \cite{bradshaw1985frank,de1976determination,hakemi1983temperature}, as well as a crystalline to nematic phase transition at 290 K for both 5CB and MBBA \cite{ahlers1994thermal,arumugam1985observation} and 263.15 K for E7 \cite{hakemi1983temperature}. In particular, the latter has been recently employed in applications to tuneable, LC integrated with metasurfaces \cite{bohn2018active}. Using the material parameters corresponding to these liquid crystals, in Figs. \ref{spontaneous_emission_5CB_fig}, \ref{spontaneous_emission_E7_fig} and \ref{spontaneous_emission_MBBA_fig} we calculate $\Gamma$ as a function of the reduced temperature $T/T_{NI}$ (normalized by the critical temperature) for all three nematics mentioned earlier for two different values of $M_{0}$ and for five different values of the critical exponent $\beta$: $\beta=0.5$ corresponding to mean-field result, $\beta=0.2$ predicted by our model [Eq.~(\ref{exponent})], $\beta=0.241$ (corresponding to the experimental work for 5CB given in \cite{chirtoc2004determination}),$\beta=0.219$ (corresponding to the experimental work for MBBA given in \cite{haller1975thermodynamic}) and $\beta=0.28$ (corresponding to the experimental work for E7 given in \cite{lenart2012tricritical}).

In the panels $(a)$ of each Figure we have set $M_{0}=1$ and they show the behavior of the spontaneous emission rate in the vicinities of the crystalline-nematic phase transition. In the crystalline phase  we define the reduced temperature $t=0.94$ for both 5CB and MBBA and $t=0.78$ for E7. 
Figures 1a, 2a and 3a reveal that the spontaneous emission rate crosses over from zero to a non-vanishing value near the crystalline-nematic phase transition, for all liquid crystals investigated. This occurs because a nonzero photon mass reduces the value of the spontaneous emission rate in the Higgs phase and eventually suppresses it for $\hbar \omega_{0} < Mc^{2}$ in which case the energy $\hbar \omega_{0}$ is smaller than the rest energy $Mc^{2}$ [see Eq. (\ref{change_in_spontaneous_emission_rate_quantum_emitter})]. This result demonstrates that one can turn on and off quantum emission in liquid crystals by varying the temperature due their critical behavior. Importantly this behavior is independent of the critical exponent $\beta$ that characterizes the phase transition.

In Figs. 1b, 2b and 3b we focus on the spontaneous emission behavior in the nematic phase only. In this case, the cross over of the spontaneous emission rate, from zero to a non-vanishing value, occurs at higher temperatures, very close to the nematic-isotropic transition temperature $T_{NI}$. Above $T_{NI}$ $\Gamma$ exhibits a monotonic increase until a certain value near the transition. For both 5CB and E7, the behavior of $\Gamma$ is very similar to each other whereas for MBBA there is a strong drop very close to the nematic-isotropic transition temperature that is not captured by the mean-field curve (dash-dotted red curve).

In addition to potential applications in the dynamical control of spontaneous emission with an external parameter, from a more fundamental point of view our findings reveal that the decay rate can probe and characterize phase transitions that occur in liquid crystals. Indeed, specially in the nematic phase, the spontaneous emission rate strongly depends on the critical exponent $\beta$, so that one could determine critical exponents at a given temperature, and hence characterize the nature of the phase transition under consideration. Again we note the good agreement between the value of the critical exponent $\beta$ analytically calculated in Eq.~(\ref{exponent}), and the experimental one given in \cite{chirtoc2004determination,lenart2012tricritical,haller1975thermodynamic} for all temperatures investigated. 

\section{\label{sec:level6} CONCLUSIONS}

In conclusion, we investigate spontaneous emission in nematic liquid crystals in the presence of an embedded quantum emitter. We develop a field theory to describe quantum emission in nematics and their critical phenomena. We discover that there exists a close analogy between this theory and the massive Stueckelberg theory, originally developed in the context of string theory. Our theory not only allows one to determine critical exponents that characterize phase transitions in liquid crystals but also to make quantitative predictions for nematics of current scientific and technological interest.  Specifically we show that the spontaneous emission rate for liquid crystals used in \cite{bohn2018active}, where they are integrated to metasurfaces, crosses over from zero to a nonvanishing value by increasing the temperature near the critical point of structural phase transitions in nematics. This finding demonstrates that one could turn on/off quantum emission in nematics as a function of temperature, allowing for unprecedented tunability and external control of quantum emission. We also predict from first principles the value of the critical exponent that characterizes the crystaline-to-nematic phase transition, which we show not only to be independent of the inertial or dissipative dynamics, but also to be in good agreement with experiments. By setting the theoretical grounds of quantum emission in liquid crystals and unveiling the role of their critical phenomena in the emission rate, we demonstrate that liquid crystals represent an efficient material platform to control and tune spontaneous emission. We hope that our findings may guide further studies on the dynamic shaping of emission spectra with liquid crystals, specially the ones where they are integrated with metasurfaces (e.g. \cite{bohn2018active}), in order to achieve dynamic, active control of quantum emission.

\begin{acknowledgments}
The authors are grateful for CAPES, CNPq, and FAPERJ for financial support. We thank V.A. Fedotov and L.S. Menezes for useful discussions.

\end{acknowledgments}

\appendix

\begin{figure}[ht]
    \centering
 	\includegraphics[width=0.4\textwidth]{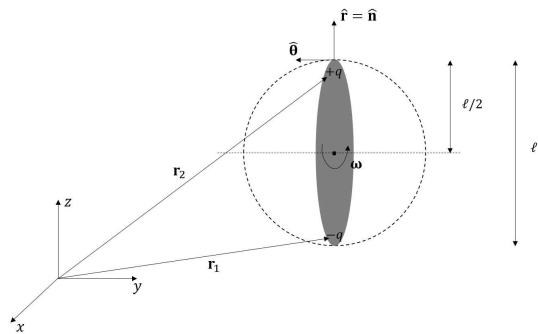}
	 \caption{Illustration of the nematic LC molecule (rod-like) as well as the coordinate systems and lengths relevant to describe it.}
	\label{nematic_molecule}
\end{figure}

\textbf{\section{\label{SingleMolHamiltonian}Derivation of the dipole Hamiltonian}}
In this appendix we present the main steps of the derivation of the Hamiltonian given in Eq.\eqref{hamiltonian_quantum_emitter}. By considering the charge distribution in the nematic molecule as described in subsection \ref{subsec2:level2}, the corresponding Lagrangian is then \cite{zangwill2013modern}
\begin{equation} \label{initial_dipole_lagragian}
    \mathcal{L}=\frac{1}{2}M(\dot{\mathbf{r}}^{2}_{1}+\dot{\mathbf{r}}^{2}_{2})+q\phi(\mathbf{r}_{1})-q\dot{\mathbf{r}}_{1}\cdot\mathbf{A}(\mathbf{r}_{1})-q\phi(\mathbf{r}_{2})+\dot{\mathbf{r}}_{2}\cdot\mathbf{A}(\mathbf{r}_{2})
\end{equation}
and with the help of Figure \ref{nematic_molecule}, that describes a rod-like LC molecule, one can show that
\begin{align}
\mathbf{r}_{1}&=\mathbf{r}_{\text{CM}}-\frac{\ell}{2}\mathbf{n}          &  \dot{\mathbf{r}}_{1}&=\dot{\mathbf{r}}_{\text{CM}}-\frac{\ell}{2}\dot{\pmb{\theta}}\times\mathbf{n} \label{charge_positions_1}\\
\mathbf{r}_{2}&=\mathbf{r}_{\text{CM}}+\frac{\ell}{2}\mathbf{n}         &  \dot{\mathbf{r}}_{2}&=\dot{\mathbf{r}}_{\text{CM}}+\frac{\ell}{2}\dot{\pmb{\theta}}\times\mathbf{n}\label{charge_positions_2}.
\end{align}
Hence, using Eqs.\eqref{charge_positions_1} and \eqref{charge_positions_2}, the Lagrangian \eqref{initial_dipole_lagragian} can be rewritten as
\begin{widetext}
\begin{multline} \label{lagrangian_dipole}
    \mathcal{L}=m\left(2\dot{\mathbf{r}}^{2}_{\text{CM}}+\frac{\ell^{2}}{2}\dot{\pmb{\theta}}^{2}\right)+q\left[\phi\left(\mathbf{r}_{\text{CM}}-\frac{\ell}{2}\mathbf{n}\right)-\phi\left(\mathbf{r}_{\text{CM}}+\frac{\ell}{2}\mathbf{n}\right)\right]+q\dot{\mathbf{r}}_{\text{CM}}\cdot\left[\mathbf{A}\left(\mathbf{r}_{\text{CM}}+\frac{\ell}{2}\mathbf{n}\right)-\mathbf{A}\left(\mathbf{r}_{\text{CM}}-\frac{\ell}{2}\mathbf{n}\right)\right]+\\
    +\frac{q\ell}{2}(\dot{\pmb{\theta}}\times\mathbf{n})\cdot\left[\mathbf{A}\left(\mathbf{r}_{\text{CM}}+\frac{\ell}{2}\mathbf{n}\right)+\mathbf{A}\left(\mathbf{r}_{\text{CM}}-\frac{\ell}{2}\mathbf{n}\right)\right],
\end{multline}
\end{widetext}
where $\mathbf{r}_{\text{CM}}$ and $m$ are the dipole's center of mass position vector and the reduced mass, respectively. 

In order to calculate the electric and magnetic potentials at the locations shown in \eqref{lagrangian_dipole}, we employ the Maxwell's equations as well as the Lorenz gauge to show first that
\begin{gather}
    \phi(\mathbf{r},t)=\phi_{0}e^{i(\mathbf{k}\cdot\mathbf{r}-\omega t+\delta_{0})} \label{electric_potential} \\
    \mathbf{A}(\mathbf{r},t)=\mathbf{A}_{0}e^{i(\mathbf{k}\cdot\mathbf{r}-\omega t+\delta_{0})} \label{magnetic_potential},
\end{gather}
with $\delta_{0}$ being a phase. Its value, however, cannot be chosen arbitrarily: we must choose $\delta_{0}=\pi/2$ in order to end up with a Stueckelberg coupling between the vector field
$\mathbf{n}$ and the radiation field $\mathbf{A}$ given by $\nabla\rightarrow\nabla+\left(e^{*}/\hbar\right)\mathbf{A}$. The absence of the imaginary unit in the coupling is required because $\mathbf{n}$ is a real field. By using \eqref{electric_potential} and \eqref{magnetic_potential}, we can rewrite \eqref{lagrangian_dipole} as
\begin{multline} \label{lagrangian_dipole_final_form}
    \mathcal{L}=m\left(2\dot{\mathbf{r}}^{2}_{\text{CM}}+\frac{\ell^{2}}{2}\dot{\pmb{\theta}}^{2}\right)-q^{*}\dot{\mathbf{r}}_{\text{CM}}\cdot\mathbf{A}(\mathbf{r}_{\text{CM}})+\\\qquad\qquad+i\wp\cos\left(\frac{\ell}{2}\mathbf{k}\cdot\mathbf{n}\right)\dot{\pmb{\theta}}\cdot\left[\mathbf{n}\times\mathbf{A}(\mathbf{r}_{\text{CM}})\right]-\\
    +q^{*}\phi(\mathbf{r}_{\text{CM}}),
\end{multline}
where $q^{*}=2q\sin\left(\frac{\ell}{2}\mathbf{k}\cdot\mathbf{n}\right)$. We have also used the triple scalar product identity from vector calculus on the second term in \eqref{lagrangian_dipole_final_form}.
The dipole Hamiltonian, hence, reads
\begin{eqnarray} \label{dipole_hamiltonian_first_form}
    \mathcal{H}=\mathbf{p}_{\text{CM}}\cdot\dot{\mathbf{r}}_{\text{CM}}+\mathbf{L}\cdot\dot{\pmb{\theta}}-\mathcal{L},
\end{eqnarray}
where the conjugate momenta $\mathbf{p}_{\text{CM}}$ and $\mathbf{L}$ are given by
\begin{gather}
    \mathbf{p}_{\text{CM}}=4m\dot{\mathbf{r}}_{\text{CM}}-q^{*}\mathbf{A}(\mathbf{r}_{\text{CM}}) \label{conjugate_linear_momentum} \\
    \mathbf{L}=m\ell^{2}\dot{\pmb{\theta}}+i\wp\cos\left(\frac{\ell}{2}\mathbf{k}\cdot\mathbf{n}\right)\left[\mathbf{n}\times\mathbf{A}(\mathbf{r}_{\text{CM}})\right]. \label{conjugate_angular_momentum}
\end{gather}
These equations lead to
\begin{gather}
    \dot{\mathbf{r}}_{\text{CM}}=\frac{1}{4m}\left[\mathbf{p}_{\text{CM}}+q^{*}\mathbf{A}(\mathbf{r}_{\text{CM}})\right] \label{velocity_vector_in_terms_of_momentum} \\
    \dot{\pmb{\theta}}=\frac{1}{m\ell^{2}}\left\{\mathbf{L}-i\wp\cos\left(\frac{\ell}{2}\mathbf{k}\cdot\mathbf{n}\right)\left[\mathbf{n}\times\mathbf{A}(\mathbf{r}_{\text{CM}})\right]\right\} \label{angular_velocity_in_terms_of_angular_momentum}
\end{gather}
 and substituting \eqref{conjugate_linear_momentum}-\eqref{angular_velocity_in_terms_of_angular_momentum} into \eqref{dipole_hamiltonian_first_form}, after a quite long but straightforward calculation, we can write down the dipole Hamiltonian in its final form:
\begin{multline} \label{dipole_hamiltonian_final_form}
    \mathcal{H}=\frac{1}{8m}\left(\mathbf{p}+q^{*}\mathbf{A}\right)^{2}-q^{*}\phi\;+\\
    +\frac{1}{2m\ell^{2}}\left[\mathbf{L}-i\wp\cos\left(\frac{\ell}{2}\mathbf{k}\cdot\mathbf{n}\right)\left(\mathbf{n}\times\mathbf{A}\right)\right]^{2},
\end{multline}
where we have omitted the mention to the center of mass position vector in order to simplify the notation.We are going to consider only the rotational degree of freedom of \eqref{dipole_hamiltonian_final_form} as indicated in \eqref{hamiltonian_quantum_emitter} and the reason for this is explained in Subsection \ref{subsec2:level2}.


\nocite{*}

\bibliography{references}

\end{document}